# Punctuated equilibrium as the default mode of evolution of large populations on fitness landscapes dominated by saddle points in the weak-mutation limit

Yuri Bakhtin[1], Mikhail I. Katsnelson[2], Yuri I. Wolf[3], Eugene V. Koonin[3,*]

[1]Courant Institute of Mathematical Sciences, New York University, 251 Mercer St, New York, NY, 10012, USA; [2]Institute for Molecules and Materials, Radboud University, Heijendaalseweg 135, NL-6525 AJ Nijmegen, Netherlands; [3]National Center for Biotechnology Information, National Library of Medicine, National Institutes of Health, Bethesda, MD 20894, USA

*For correspondence: koonin@ncbi.nlm.nih.gov

## Abstract

Punctuated equilibrium is a mode of evolution in which phenetic change occurs in rapid bursts that are separated by much longer intervals of stasis during which mutations accumulate but no major phenotypic change occurs. Punctuated equilibrium has been originally proposed within the framework of paleobiology, to explain the lack of transitional forms that is typical of the fossil record. Theoretically, punctuated equilibrium has been linked to self-organized criticality (SOC), a model in which the size of 'avalanches' in an evolving system is power-law distributed, resulting in increasing rarity of major events. We show here that, under the weak-mutation limit, a large population would spend most of the time in stasis in the vicinity of saddle points in the fitness landscape. The periods of stasis are punctuated by fast transitions, in $\ln N_e$ time ($N_e$, effective population size), when a new beneficial mutation is fixed in the evolving population, which moves to a different saddle, or on much rarer occasions, from a saddle to a local peak. Thus, punctuated equilibrium is the default mode of evolution under a simple model that does not involve SOC or other special conditions.

## Significance

The gradual character of evolution is a key feature of the Darwinian worldview. However, macroevolutionary events are often thought to occur in a non-gradualist manner, in a regime known as punctuated equilibrium, whereby extended periods of evolutionary stasis are punctuated by rapid transitions between states. Here we analyze a mathematical model of population evolution on fitness landscapes and show that, for a large population in the weak-mutation limit, the process of adaptive evolution consists of extended periods of stasis, which the population spends around saddle points on the landscape, interrupted by rapid transitions to new saddle points when a beneficial mutation is fixed. Thus, punctuated equilibrium appears to be the default regime of biological evolution.

## Introduction

Phyletic gradualism, that is, evolution occurring via a succession of mutations with infinitesimally small fitness effects, is a central tenet of Darwin's theory (1). However, the validity of gradualism has been questioned already by Darwin's early, fervent adept, T.H. Huxley (2), and subsequently, many non-gradualist ideas and models have been proposed, to account, primarily, for macroevolution. Thus, Goldschmidt (in)famously championed the hypothesis of "hopeful monsters", macromutations that would be deleterious in a stable environment but might give their carriers a chance for survival after a major environmental change (3). Arguably, the strongest motivation behind non-gradualist evolution concepts was the notorious paucity of intermediate forms in the fossil record. It is typical in paleontology that a species persists without any major change for millions of years, but then, is abruptly replaced by a new one. The massive body of such observations prompted Simpson, one of the founding fathers of the Modern Synthesis of evolutionary biology, to develop the concept of quantum evolution (4), according to which species, and especially, higher taxa emerged abruptly, in 'quantum leaps', when an evolving population rapidly moves to a new 'adaptive zone', or using the language of mathematical population genetics, a new peak on the fitness landscape. Simpson proposed that the quantum evolution mechanism involved fixation of unusual allele combinations in a small population by genetic drift, followed by selection driving the population to the new peak.

The idea of quantum evolution received a more systematic development in the concept of punctuated equilibrium (PE) proposed by Eldredge and Gould (5-8). The abrupt appearance of species in the fossil record prompted Eldredge and Gould to postulate that evolving populations of any species spend most of the time in the state of stasis, in which no major phenotypic changes occur (9, 10). The long intervals of stasis are punctuated by short periods of rapid evolution during which speciation occurs, and the previous dominant species is replaced by a new one. Gould and Eldredge emphasized that PE was not equivalent to the "hopeful monsters" idea, in that no macromutation or saltation was proposed to occur, but rather, a major acceleration of evolution via rapid succession of 'regular' mutations that resulted in the appearance of instantaneous speciation, on geological scale.

A distinct but related view of macroevolution is encapsulated in the concept of evolutionary transitions developed by Szathmary and Maynard Smith (11-13). Under this concept, major evolutionary transitions, such as, for instance, emergence of multicellular organisms, involve emergence of new levels of selection (new Darwinian individuals), in this case, selection affecting ensembles of multiple cells rather than individual cells. These evolutionary transitions resemble phase transitions in physics (14)and appear to occur rapidly, compared to the intervals of evolution within the same level of selection. The concept of evolutionary transitions can be generalized to apply to the emergence of any complex feature (15).

Punctuated equilibrium has been explicitly linked to the physical theory of self-organized criticality (SOC). Self-organized criticality, a concept developed by Bak and colleagues (16), is an intrinsic property of dynamical systems with multiple degrees of freedom and strong nonlinearity. Such systems experience serial 'avalanches' separated in time by intervals of stability (the avalanche metaphor comes from Bak's depiction of SOC on the toy example of a sand pile, on which additional sand is poured, but generally denotes major changes in a system). A distinctive feature of the critical dynamics under the SOC concept is self-similar (power law) scaling of avalanche sizes (16-22). The close analogy between SOC and PE was noticed and explored by Bak and colleagues, the originators of the SOC concept, who developed models directly inspired by evolving biological systems and intended to describe their behavior (16, 19, 20, 22). In particular, the popular Bak-Sneppen model (19) explores how ecological connections between organisms (physical proximity in the model space) drive co-evolution of the entire community. Extinction of the organisms with the lowest fitness disrupts the local environments and results in concomitant extinction of their closest neighbors. It has been shown that, after a short burn-in, such systems self-organize in a critical quasi-equilibrium interrupted by avalanches of extinction, with the power law distribution of avalanche sizes.

We asked whether SOC is a prerequisite for PE and, more broadly, what are the necessary and sufficient conditions for PE. To address this question, we analyze mathematically a simple model of population evolution on a rugged fitness landscape (23). We show that, under the assumptions of a large population size and low mutation rate (weak-mutation limit), an evolving population spends most of the time in stasis, i.e. percolating in a near-neutral mutational networks around saddle points on the landscape. The intervals of stasis are punctuated by rapid transitions to new saddle points after fixation of beneficial mutations. Thus, contrary to the general perception of the weak-mutation limit as an equivalent of gradualism (24), PE appears to be the default mode of evolution of large populations in this regime.

## Results
### *Agent-based model of competitive exclusion*

We consider a population of a large constant size $N$ consisting of individuals, each with a specific genotype. To avoid dealing with the overwhelming complexity of the space of all genotypes, we work with a coarse-grained model that groups similar genotypes into 'types'. The genotypes within the same type are considered to be homogeneous and densely connected by the mutation network. The only homogeneity assumption we need to make is that, within each type, the variations in fitness and available transitions to other classes due to mutations are negligible. We also assume that sizes of different types are comparable. The set of all types is denoted by $\mathbb{T}$.

The evolution of a population within the model involves reproduction and mutation. Reproduction of individuals occurs under the Moran model widely used in population genetics, that is, with rates proportional to their fitness and is accompanied by removal of random individuals to keep $N$ constant (25). Mutations are modeled by transitions in a mutational network $E$. The individual mutation rate $\lambda$ is assumed to be low compared to the reproduction rates. The evolutionary regime depends on: i) the geometry of the graph ($\mathbb{T}$, $E$), ii) the fitness function $f$, iii) the values of parameters $N$ and $\lambda$, iv) the initial configuration.

Let us now describe our basic model in more detail. We assume that the population size is a large number $N$, constant in time. The set $\mathbb{T}$ of all possible types is finite or countable. It can be viewed as a graph with adjacency matrix $(E_{ij})_{i,j \in \mathbb{T}}$. Two distinct types $i, j$ are connected by an edge if they differ by a mutation (at the scale of the model, a mutation is assumed to occur instantaneously and without intermediate steps). In that case, we set $E_{ij} = 1$. Otherwise, $E_{ij} = 0$.

Each type $i \in \mathbb{T}$ is assigned a fitness value $f_i > 0$ which is identified with the reproduction rate. The numbers $f_i$ are assumed to be distinct and of the order of 1 (more precisely, bounded), so essentially, time is measured in reproductions. It is convenient to work with relative sizes $y_i$ of type populations (fractions) with respect to the total population size $N$. We denote by $\Delta$ the space of sequences $(y_i)_{i \in \mathbb{T}}$ such that $y_i \geq 0$ for all $i$ and $\sum_{i \in \mathbb{T}} y_i = 1$. Denoting the fraction of individuals of type $i \in \mathbb{T}$ present in the population at time $t \in \mathbb{R}$ by $x_i(t)$ (taking values $0, N^{-1}, 2N^{-1}, ...$), we define random evolution of the vector $(x_i(t))_{i \in \mathbb{T}} \in \Delta$ as a continuous time pure jump $\Delta$-valued Markov process, by specifying the transition rates. A single individual of type $i \in \mathbb{T}$ produces new individuals of the same type $i$ at the rate $f_i$. Each reproduction is accompanied by removal of one individual that is randomly and uniformly chosen from the entire population. Thus, the total rate of reproduction of individuals of type $i$ is $N x_i f_i$. Given that an individual of type $i$ is reproducing, the probability that the child individual will replace an individual of type $j$ is $x_j$. Thus, the total rate of simultaneous change $x_i \rightarrow x_i + N^{-1}$ and $x_j \rightarrow x_j - N^{-1}$ is $Nf_i x_i x_j$. Let us now introduce mutations. We will assume that mutation rates are much lower than the reproduction rates. To model this, we introduce a small parameter $\lambda > 0$. The rate of replacement of an individual of type $i \in I(x)$, where

$$I(x) = \{i \in \mathbb{T} : x_i > 0\}, \quad x \in \Delta,$$

by an individual of type $j$ is given by $\lambda E_{ij} \in \{0, \lambda\}$. The total rate of such transitions occurring in a population is $N \lambda E_{ij} x_i$.

In what follows, we derive the PE evolutionary regime from certain reasonable assumptions on the geometry of the graph, the fitness function, population size, mutation rates and the initial state. Our results can be viewed as similar to those in previous work (26-28), where more sophisticated models were considered. However, our simple model allows for a more transparent analysis that is conducive to biological implications and we use it here to tie the PE concept to noisy dynamics near heteroclinic networks (29, 30) and emphasize the importance of saddle points on the landscape for the evolutionary process.

### *Evolution without mutations in the infinite population size limit*

In this section, we examine the case where, in an infinite population, $\lambda = 0$, i.e., there are no mutations, and approximate the dynamics of our stochastic model by that of a deterministic ODE

$$\dot{x}_i = b_i(x), \quad i \in \mathbb{T}, \tag{1}$$

with the right-hand side given by

$$b_i(x) = x_i(f_i - \bar{f}(x)),$$

where $\bar{f}(x) = \sum_{i \in \mathbb{T}} f_j x_j$ is the average fitness for the population state $x$. The system (1) is a well-known competitive exclusion system (see, e.g., (2.15)–(2.16) of (31)) restricted to nonzero components of $x$. Equation (1) emerges due to the averaging effect and can be viewed as a law of large numbers for our model.

To state our results, we need to introduce some notations and definitions. We denote $I = I(x(0))$ for brevity and note that, given the absence of mutations, our stochastic model and ODE (1) are defined on the simplex $\Delta_I = \{x \in \mathbb{R}_+^I : \sum_{i \in I} x_i = 1\}$. This simplex is the convex hull of its vertices $e^{(i)}$, $i \in I$, corresponding to pure states where only one type is present:

$$e_k^{(i)} = \begin{cases} 1, & i = k, \\ 0, & i \neq k. \end{cases}$$

One of these vertices plays a special role. Let $i^*$ be the type with maximum fitness $f^*$ (within $I$), that is, $f^* = f_{i^*} = \max_{i \in I} f_i$. We will see that $e^{(i^*)}$ is an attractor for both deterministic dynamical system defined by (1) and for our stochastic model. For the approximation result, we need to define the discrepancy

$$D(t) = x(t) - \Phi^t x(0), \tag{2}$$

where $x(t)$ is the Markov process without mutations and for any $y$, $\Phi^t y$ is the solution of ODE (1) with the initial condition $y$, at time $t$. We are going to estimate the maximal discrepancy up to time $t$, i.e., $D^*(t) = \sup_{s \in [0,t]} \| D(s) \|$, where $\|\cdot\|$ is the $L^1$ norm in $\mathbb{R}^I$ defined by

$$\| x \| = \sum_{i \in I} | x_i |. \tag{3}$$

We assume that the number of types $|I|$ is small compared to the population size, more precisely, there is $\mu < 1/2$ such that

$$|I| \leq N^\mu. \tag{4}$$

Because this model does not include mutations, if a type $i$ becomes extinct at time $s$, i.e., $x_i(s) = 0$, then, $x_i(t) = 0$ for all $t \geq s$. We denote the event on which no type $i \in I$ becomes extinct before time $t$ by $B_t = \{I(x(s)) = I \text{ for all } s \in [0, t]\}$. Events from a sequence $(A_N)_{N \in \mathbb{N}}$ are stretch-exponentially unlikely (SE-unlikely) if for some $C, \gamma > 0$,

$$\mathsf{P}(A_N) \leq C e^{-N^\gamma}, \quad N \in \mathbb{N}.$$

This is fast decay in $N$, just short of being truly exponentially fast. We are now ready to state our main result for the system without mutations and to examine on the meaning of each of its parts.

**Theorem 1**. *Assume (4). Then:*

1. *There are constants $c, \beta > 0$ such that events $B_{c \ln N} \cap \{D^*(c \ln N) > N^{-\beta}\}$ are SE-unlikely.*

2. *Let $\beta$ be defined in Part 1 of the Theorem. Then, for any $\delta < \beta$, there is a constant $C > 0$ such that, conditioned on the nonextinction of type $i^*$, and up to a SE-unlikely event, $|x(C \ln N) - e^{(i^*)}| \leq N^{-\delta}$.*

3. *There are constants $C', \alpha > 0$ such that, if $|x(0) - e^{(i^*)}| \leq N^{-\delta}$, then*

$$\mathsf{P}\big\{x(C'\mathrm{ln}N) = e^{(i^*)}\big\} > 1 - N^{-\alpha}.$$

4. *There is a number $p > 0$ that does not depend on N, such that the probability of nonextinction of type $i^*$ is bounded below by $p$ for all initial conditions $x(0)$ satisfying $x_{i^*}(0) > 0$.*

5. *For any $\delta \in (0,1)$, if $x_{i^*}(0) > N^{-\delta}$, then, extinction of type $i^*$ is SE-unlikely.*

Part 1 of the theorem shows that, up to time $c\mathrm{ln}N$, if no type gets extinct, the stochastic process $x(t)$ follows the deterministic trajectory $\Phi^t x(0)$ very closely, deviating from it at most by $N^{-\beta}$. This happens with a probability very close to 1, exceptions being stretch-exponentially unlikely.

Part 2 shows that, if type $i^*$ does not die out, then, with high probability, by time $C\mathrm{ln}N$, it will dominate the population and all other types will be almost extinct.

Part 3 means that, after realization of the scenario described in Part 2 and an additional logarithmic time, $i^*$ will be the only surviving type.

Part 1 is conditioned on the nonextinction of any type, whereas Part 2 is conditioned on the nonextinction of type $i^*$. If any type $i$ dies out, Part 1 still applies to the continuation of the process on the simplex $\Delta_{I\setminus\{i\}}$ of a lower dimension. By contrast, for Part 2 to be meaningful, we need to provide a bound on the nonextinction of $i^*$. This is done in Parts 4 and 5.

Part 4 states that there is a positive probability (independent of the population size) that the progeny of even a single individual of type $i^*$ will drive out all other types.

Part 5 states that, once the fraction of the individuals of type $i^*$ reaches a (small) threshold $N^{-\delta}$, then, it is almost certain that $i^*$ will dominate the population. To summarize these results, the chance of extinction for the fittest type is non-negligible only when there are very few individuals of this type, that is, when the initial state involves a recent mutation that produced a single individual of this type. Once the number of individuals reaches a certain modest threshold, the typical, effectively deterministic, behavior is to follow the trajectory of (1) closely, eventually reaching the pure state of fixation where only individuals of type $i^*$ are present. The proof of Theorem 1 is given in the Appendix. Now, we turn to the analysis of the dynamics generated by ODE (1).

### *Behavior of the deterministic system*

In this section, we explore the behavior of the system (1). Our basic analysis is only a minor extension of previous work (31)(Section 2.2.1), and we include it here for completeness and to stress the points central to the concept of evolution in the PE regime that is developed in this paper. The first statement characterizes the survival of the fittest under this dynamic.

**Theorem 2.** *Let $x(t)$ be a solution of Eq. (1). If $x_{i^*}(0) > 0$, then $x(t)$ converges to $e^{(i^*)}$ exponentially fast.*

One possible approach to the proof of this theorem is to define

$$\tilde{f} = \max_{i \in I \setminus \{i^*\}} f_i < f^*, \tag{5}$$

and note that

$$\bar{f}(t) \leq x_{i^*}(t) f^* + (1 - x_{i^*}(t)) \tilde{f}$$

together with equation (1) implies

$$\dot{x}_{i^*}(t) \geq x_{i^*}(t)(f^* - x_{i^*}(t)f^* - (1 - x_{i^*}(t))\tilde{f}) = x_{i^*}(t)(1 - x_{i^*}(t))(f^* - \tilde{f}),$$

Therefore, $y(t) = 1 - x_{i^*}(t)$ satisfies

$$\dot{y}(t) \leq -c(1 - y(t))y(t),$$

where $c = f^* - \tilde{f} > 0$. Thus, $y(t)$ is dominated by the solution of the equation $\dot{z} = -c(1-z)z$ which converges to zero exponentially fast, so $1 - x_{i^*}(t) \leq Ke^{-ct}$ for some $K > 0$ depending on the initial condition, which completes the proof.

Here, our assumption that $f$ takes distinct values was used to ensure that the constant $c$, the gap between the maximum value of $f$ and the second highest value (this constant also plays the role of the convergence rate), is positive. If the maximum fitness is attained by several distinct types (as opposed to essentially indistinguishable microstates within a type), then, a similar estimate shows that, in the limit, only those maximum fitness types survive.

Although the analysis above already allows us to conclude that points $e^{(k)}$ are hyperbolic critical points (saddles) of various indices (the index of a saddle is the number of negative eigenvalues of the linearization of the vector field at the saddle), we can show this more explicitly. It is easy to compute the linearization $(\partial_j b_i(e^{(k)}))$ of $b$ at $e^{(k)}$:

$$\begin{aligned}
\partial_k b_k(e^{(k)}) &= -f_k, \\
\partial_i b_k(e^{(k)}) &= -f_i, && i \neq k, \\
\partial_i b_i(e^{(k)}) &= f_i - f_k, && i \neq k, \\
\partial_j b_i(e^{(k)}) &= 0, && j \neq i, i \neq k.
\end{aligned}$$

Therefore, for each $i \in I$ such that $i \neq k$, there is an eigenvalue $f_i - f_k$ of $(\partial_j b_i(e^{(k)}))$ with an eigenvector $e^{(i)} - e^{(k)}$ pointing along the simplex edge connecting $e^{(k)}$ and $e^{(i)}$. These eigenvalues span the simplex $\Delta_I$, so the additional eigenvalue $-f_k$ with eigenvector $e^{(k)}$ that is transversal to $\Delta_I$ can be ignored. To demonstrate explicitly that the vertex $e^{(k)}$ is a saddle, we note that the eigendirections given by $e^{(i)} - e^{(k)}$ are stable or unstable, depending on the sign of the associated eigenvalue, i.e., on whether $f_i < f_k$ or $f_i > f_k$. Moreover, there is a heteroclinic connection (a trajectory connecting two distinct saddle points) between $e^{(i)}$ and $e^{(k)}$. This trajectory coincides with the simplex edge between $e^{(i)}$ and $e^{(k)}$ and corresponds to the presence of exactly two types $i, k$. The dynamics on it is described by the logistic equation

$$\dot{x}_i = (f_i - f_k)x_i(1 - x_i).$$

(see Figure 1 for the phase portrait). The key feature of this dynamics is a heteroclinic network formed by trajectories connecting saddle points to one another. The vertex $e^{(i^*)}$ is a sink (a saddle of index 0) if considered in $\Delta_I$ but it can also be viewed as a saddle in simplices of higher dimensions based on coordinates (types) that include those with higher fitness than $f^*$. The types with higher fitness will appear if we include mutations into the model.

### *Evolutionary process with mutations*

We now consider the full process with positive but small rate $\lambda$ and recall that, for each type $i \in I(x)$, the rate of mutation to type $j$ is given by $\lambda E_{ij}$. We consider here only relatively late stages of evolution that are preceded by extensive evolutionary optimization so that the overwhelming majority of the mutations are either deleterious or at best neutral. More precisely, we assume that there is a constant $M$ such that for each $i \in I(x)$, the total number of available fitness-increasing (beneficial) mutations, that is, vertices $j \in \mathbb{T}$ such that $E_{ij} = 1$ and $f_j > f^*$, is bounded by $M$. Our first assumption on the magnitude of $\lambda$ is that

$$r(N) = \lambda N \ln N \ll 1.$$

Then, for a fixed $C > 0$, large $N$, and any time interval of length $C\ln N$, the probability of a beneficial mutation is bounded by

$$1 - e^{-MN\lambda C\ln N} = 1 - e^{-MCr(N)} \leq MCr(N). \tag{6}$$

According to Theorem 1, if the evolutionary process is conditioned on the survival of type $i^*$, then, typically, it takes $C\ln N$ time for the process $x_{i^*}(t)$ to reach 1 (fixation). Thus, the estimate (6) shows that the population is unlikely to produce a new beneficial mutation before it reaches the state of fixation where type $i^*$ is the only surviving one. Once a new beneficial mutation occurs and, accordingly, a new best-fit type emerges, it either gets extinct quickly or gets fixed in the population, in time of the order $\ln N$. The trajectory, driven by differential reproduction of random mutations, closely follows the heteroclinic connection, i.e., the line connecting two vertices of the simplex $\Delta$. The entire process can be described as follows: there is a moment when $i^*$ is the only type present, after which it takes time of order $(k\lambda N)^{-1}$ to produce a new beneficial mutation, where $k$ is the number of beneficial mutations that are available from $i^*$. Then, it takes a much shorter time $C\ln N$ for this fittest type to take over the entire population, after which the process repeats.

Now consider deleterious mutations. There are $N$ individuals, and each produces a suboptimal (lower fitness) type with the rate $\lambda L$, where $L$ is the number of available deleterious mutations. Using the Poisson distribution, we obtain that, by time $t$, it is highly unlikely to produce more than $tN\lambda L$ new suboptimal individuals. If $t = C \log N$, then, this number is $C\lambda LN \ln N$, so requiring

$$\lambda L\ln N \ll 1, \tag{7}$$

we obtain $\lambda LN\ln N \ll N$, that is, over the travel time between saddles, the emerging individuals with deleterious mutations constitute an asymptotically negligible fraction of the entire population. Thus, the trajectory $x(t)$ will be altered only by a term converging to 0 as $N \to \infty$.

Thus, the emerging picture is as follows: the evolving population spends most of the time in a 'dynamic stasis' near saddle points. During this stage, a dynamic equilibrium emerges under purifying selection: deleterious mutations constantly produce individuals with fitness lower than the current maximum, and these individuals or their progeny die out. On time scale of $(k\lambda N)^{-1}$, a new beneficial mutation will occur, and then, either the new type will go extinct fast (in which case, the population has to wait for another beneficial mutation) or will get fixed such that, in time $\ln N$, the new type (followed by a small, dynamic cloud of suboptimal types) will dominate the population. The transition from one dominant type to the next occurs along the heteroclinic trajectory ~~orbit~~ coinciding with the edge of the infinite-dimensional simplex connecting the two vertices corresponding to monotypic populations. This iterative process of fast transitions between long stasis periods spent near saddle points is typical of noisy heteroclinic networks, as demonstrated in early, semi-heuristic work (32) (33, 34), and later, rigorously(29, 30). However, the two types of noisy contributions, from reproduction and mutation, play distinct roles here, so although the general punctuated character of the process that we describe here is the same as in the previous studies, their results do not apply to our case straightforwardly.

Because the process is random, deviations from this general description eventually will occur. Stretch-exponentially unlikely, extremely rare events can be ignored. However, the right-hand side of Eq. (6), albeit small, does not decay stretch-exponentially, and so, with a non-negligible frequency, a new beneficial mutation would appear before the current fittest type takes over the entire population. The result will be clonal interference such that the current fittest type starts being replaced with the new one before reaching fixation.

### *Taking the structure of the landscape into account*

In general, the structure of the landscape can be complicated. The available information on the structure of complex landscapes is limited, and there are few mathematical results. Several rigorous results based on random matrix theory have been obtained for centered Gaussian fields on Euclidean spheres of growing dimension with rotationally invariant covariances of polynomial type (35, 36). For those models, the average numbers of saddles of different indices at various levels of the landscape have been shown to grow exponentially with respect to the dimension of the model, and a variational characterization of the exponential rates has been obtained. Although formally limited to concrete models, these results indicate that there are many local maxima and many more saddle points in such complex landscapes. In the context of the evolutionary process, this indicates that the evolutionary path through a sequence of temporarily dominant types is likely to end up not in a global but in a local maximum. Consider now what transpires near a local fitness peak. Suppose the current dominant genotype differs in $k_0$ sites from the locally optimal genotype, and sequential beneficial mutations in these sites in an arbitrary order produce a succession of increasing fitness values. Ignoring shorter times of order $\ln N$ of transitioning between saddles and only taking into account the leading contributions (that is, the sum of the waiting times for the beneficial mutations), the time it takes to reach the peak is then of the order of $(k_0\lambda N)^{-1} + ((k_0 - 1)\lambda N)^{-1} + \cdots + (2\lambda N)^{-1} + (\lambda N)^{-1} \approx (\lambda N)^{-1}\ln k_0$ (recall that our time units are comparable with reproduction rates). Once the peak is reached, it is extremely unlikely that the population moves anywhere else on the landscape. More specifically, the waiting time for the appearance of a new dominant genotype is exponentially large in $N$ as follows from the metastability theory at the level of large deviations estimates.

## Discussion

Fossil record analysis suggests that PE dominates organismal evolution (7, 8, 10). Here we examine mathematically a simple population-genetic model and show that PE is the default regime of population evolution under basic, realistic assumptions, namely, large effective population size, low mutation rate and rarity of beneficial mutations. In the weak-mutation limit, large populations spend most of their time in 'dynamic stasis', i.e. exercising short-range random walks within their local neutral networks, without shifting to a new distinct state in the vicinity of saddle points on the fitness landscape. The stasis periods are punctuated by rapid transitions between saddle points upon emergence of new beneficial mutations; these transitions appear effectively instantaneous compared to the duration of stasis (Figure 2). Eventually, the population might reach a local fitness peak where no beneficial mutations are available. This would lead to indefinite stasis as long as the fitness landscape does not change and the population size stays large (drift to a different peak is exponentially rare in $N_e$, that is, impractical for large $N_e$).

Two conditions determine the behavior described by this model: i) smallness of the overall mutation rate (dominated by the deleterious mutations), eq (7), $\lambda L \ll 1/\ln N$ and ii) smallness of the beneficial mutation rate, which results in the difference in scale between the waiting time $(\lambda k N)^{-1}$ and the saddle-to-saddle transition time $\ln N$, i.e. $\lambda k N \ll 1/\ln N$. Comparison of the expressions for these conditions suggests that, for the PE to be pronounced, deleterious mutations should outnumber the beneficial mutations by at least a factor of $N$. This is a large but not unrealistic difference in the case of 'highly adapted' organisms, that is, in situations, most common in the extant biosphere, where the pool of trivial optimizations that presumably were available at the earlies stages of the evolution of life, is exhausted.

For example, with population and genomic parameters characteristic of animals, $N$ of ~$10^5$ and ~$10^7$ amino acid-encoding sites in the genome, the local mutational neighborhood in the sequence space consists of $19 \times 10^7$ mutations. Assuming that about half of these mutations are deleterious and noting that the number of beneficial mutations should be less by a factor of $10^5$, there must be $1<k<1000$ beneficial mutations, apparently, a realistic value.

The condition on the overall mutation rate ($\lambda L \ll 1/\ln N$) is more difficult to assert because both $\lambda$ and $L$ depend on the clustering of the whole sequence space into a coarse-grained network of distinct types. Note, however, that, as the first approximation, $\lambda$ is bounded by the sequence-level mutation rate $\mu$ (only some of the sequence-level mutations lead to transitions between distinct types) and $L$ is bounded by the genome size $G$ (the number of available sequence-level single-position mutations is on the order of the genome size, but only some of these mutations have detectable deleterious effect). Thus, $\lambda L < \mu G$, where $\mu G$ is the expected number of sequence-level mutations per genome per generation. It has been shown that the values of $\mu G$ tend to stay of the order of $1/N$ under 'normal' conditions (37, 38), therefore

$$\lambda L < \mu G \sim 1/N \ll 1/\ln N$$

so that the weak-mutation regime is likely to hold under broad range of conditions.

Thus, our model suggests that the PE regime is common in the evolution of natural populations. The probable exceptions include stress-induced mutagenesis (39), whereby the mutation rate can rise by orders of magnitude, locally blooming microbial populations that might violate the $kN \ll L$ condition, and abrupt changes in the fitness landscape that might temporarily increase the number of immediately beneficial mutations $k$. All of these situations, however, are likely to be transient.

Theoretically, PE has been linked to SOC as the underlying mechanism (16, 19). However, we show here that PE naturally emerges in extremely simple models of population evolution that do not involve any criticality. The major conclusion from this analysis is that PE and not gradualism is the fundamental characteristic of sufficiently large populations in the weak-mutation limit which is, arguably, the most common evolutionary regime across the entire diversity of life. The parameter values that lead to PE appear to hold for evolving populations of all organisms, including viruses, under 'normal' conditions. Situations can emerge in the course of evolution when the PE regime breaks through disruption of the stasis phase. This could be the case in very small populations that rapidly evolve via drift or in cases of a dramatically increased mutation rate, such as stress-induced mutagenesis, and especially, when these two conditions combine (39-41). In many cases, disruption of stasis will lead to extinction but, on occasion, a population could move to a different part of the landscape, potentially, the basin of attraction of a higher peak. The evolution of cancers, at least, at advanced stages, does not appear to include stasis either, due to the high rate of nearly neutral and deleterious mutations, and low effective population size (39). Furthermore, the PE regime is characteristic of 'normal' evolution of well-adapted populations in which the fraction of beneficial mutations is small. If many, perhaps, the majority of the mutations are beneficial, there will be no stasis but rather a succession of rapid transitions in a fast adaptive evolution regime. Conceivably, this was the mode of evolution of primordial replicators at pre-cellular stages of evolution.

One of the most fundamental – and most difficult – problems in biology is the origin of major biological innovations (more or less, synonymous to macroevolution). In modern evolutionary biology, Darwin's central idea of survival of the fittest transformed into the concept of fitness landscape with numerous peaks, where each stable form occupies one of the peaks (23, 42). Then, the fundamental problem arises: if a population has reached a local peak, further adaptive evolution is possible only via a stage of temporary decrease of fitness – how can this happen? A common answer is based on Wright's concept of random genetic drift: the smaller the effective population size $N_e$, the greater the probability of random drift through (not excessively deep) valleys in the fitness landscape (42-44). This notion implies that major evolutionary transitions occur through narrow population bottlenecks. As formalized in our previous work, the evolutionary 'innovation potential' is inversely proportional to $N_e$ (14). There are, however, multiple indications that drift cannot be the only mode of evolutionary innovation and that novelty often arises in large populations thanks to their high mutational diversity (45-48). Nevertheless, it remains unclear, within the tenets of classical population genetics, how a large population can cross a valley on the landscape. One obvious way to overcome this conundrum is to assume that the landscape changes in time due to environmental changes, so that a population can find itself in the basin of attraction of a new fitness peak (49, 50).

The analysis presented here suggests a greater innovation potential of large populations than usually assumed, stemming from the fact that a typical landscape in a multidimensional space contains many more saddle points than peaks. On the one hand, this intuitively obvious claim follows from the observation that, for any two peaks, the path connecting the peaks and maximizing the minimum height must pass through a saddle point. On the other hand, it is justified by precise computations of exponential (with respect to the model dimension) growth rates of the expected numbers of saddle points of various indices (including peaks) for random Gaussian landscapes under certain restrictions on covariance (35, 36). Thus, typical fitness landscapes are likely to allow numerous transitions and extensive, innovative evolution without the need for valley crossing.

In biological terms, it seems to be impossible to maximize fitness in all numerous directions (the number of these being at least on the order of the genome size), and therefore, the probability of beneficial mutations is (almost) never zero, however small it might be (in general, this pertains not only to single point mutations, but also to beneficial epistatic combinations of mutations as well as large scale genomic changes, such as gene gain, loss and duplication). In other words, the landscape is dominated by saddle points that are far more common than peaks, so that there is almost always an upward path which an evolving population will follow provided it is large enough to afford a long wait in saddles without risking extinction due to fluctuations.

Results similar to ours have been reported in the mathematical biology literature (26-28). Specifically, it has been proven that a trait substitution sequence process (sequential transition from one dominant trait to another) occurs in the limit of large population size and small beneficial mutation rate. Here we employ a very simple model to demonstrate the fundamental character of the concept of punctuated equilibrium, to tie it to the noisy dynamics near heteroclinic networks (29, 30) and to stress the key role of saddle points, in contrast to the wide-spread perception of peaks as the central structural elements of fitness landscapes.

To conclude, the results presented here show that PE is not only characteristic of speciation or evolutionary transitions but rather is the default mode of evolution under weak-mutation limit which is the most common evolutionary regime (24). In our previous work, we have identified conditions under which saltational evolution becomes feasible, under the strong-mutation limit (41). Here we show that, even for evolution in the weak-mutation limit that is generally perceived as gradual (24), PE is the default regime. Even during periods of stasis in phenotypic evolution, the underlying microevolutionary process appears to be punctuated.

## Author contributions

YB, MIK, YIW, and EVK jointly incepted the project; YB performed the mathematical analysis; YB, MIK, YIW, and EVK analyzed the results; YB and EVK wrote the manuscript that was edited and approved by all authors.

## Acknowledgements

YIW and EVK are supported by the Intramural Research Program of the National Institutes of Health of the USA. YB is partially supported by the National Science Foundation, grant DMS-1811444. MIK was supported by Spinoza Prize funds.

Figure legends

Figure 1. **The phase portrait of the dynamical system (1).**

Four types 1, 2, 3, 4 are shown such that $f_1 < f_2 < f_3 < f_4$. The dynamics is defined on the simplex $\Delta_{\{1,2,3,4\}}$ with vertices $e^{(1)}, e^{(2)}, e^{(3)}, e^{(4)}$, corresponding to pure states where the population consists entirely of individuals of one type. These vertices are critical points of the vector field $b$. The edges of the simplex are heteroclinic orbits connecting these critical points to each other. Several other orbits are also plotted as arrows. The vertex $e^{(4)}$ attracts every initial condition with nonzero fraction of individuals of the fittest type $i^* = 4$.

Figure 2. **Evolution under punctuated equilibrium on a fitness landscape dominated by saddles: stasis around saddle points punctuated by fast adaptive transitions.**

Planar shapes depict distinct classes of genotypes. The color scale shows a range of fitness values. Gray "ramp" strips show available transitions between the genotype classes ($k$ transitions leading to classes with higher fitness and $L$ transitions leading to classes with lower fitness, $k \ll L$). The two blue circles indicate the original and the current states of the population; blue arrows show succession of genotypes within the same class, occurring within the effectively neutral network during the "dynamic stasis" phase; red arrows indicate fast adaptive transitions from a lower-fitness genotype to one with a higher fitness.

## Appendix

### *Proof of Theorem 1*

To prove Part 1, our first goal is to represent the discrepancy $D(t)$ defined in (2) in a convenient way. We can write the solution $\Phi^t x(0)$ of ODE (1) with initial value $x(0)$ as

$$(\Phi^t x(0))_i - x_i(0) = \int_0^t b_i (\Phi^s x(0))ds, \quad i \in I. \tag{8}$$

It is useful to represent $x(t)$ in a similar form. To that end, we recall that every Markov process solves the martingale problem associated with its own generator. Therefore, introducing the projection function $\pi_i(x) = x_i$, we obtain that there is a martingale $M_i$ such that

$$x_i(t) - x_i(0) = \pi_i(x(t)) - \pi_i(x(0)) = \int_0^t \mathcal{N}\pi_i(x(s))ds + M_i(t), \quad i \in I, \tag{9}$$

where the generator $\mathcal{N}h$ is defined by

$$\mathcal{N}h(x) = \lim_{t \downarrow 0} \frac{\mathsf{E}[h(x(t))|x(0) = x] - h(x)}{t}.$$

For our pure jump process the generator is determined by transition rates:

$$\mathcal{N}h(x) = N \sum_{\substack{i,j \in I \\ i \neq j}} f_i\, x_i x_j (h(\sigma^{ij} x) - h(x)),$$

where $\sigma^{ij} x$ denotes the state obtained from state $x$ by adding an individual of type $i$ displacing an individual of type $j$:

$$(\sigma^{ij} x)_k = \begin{cases} x_k, & k \neq i, j, \\ x_i + \dfrac{1}{N}, & k = i, \\ x_j - \dfrac{1}{N}, & k = j. \end{cases}$$

We can compute directly:

$$\mathcal{N}\pi_i(x) \quad = N \sum_{j:j \neq i} f_i\, x_i x_j \frac{1}{N} + N \sum_{j:j \neq i} f_j\, x_j x_i \left(-\frac{1}{N}\right) = \sum_j (f_i - f_j)x_i x_j = b_i(x).$$

Plugging this into (9), we obtain

$$x_i(t) - x_i(0) = \int_0^t b_i (x(s))ds + M_i(t), \quad i \in I. \tag{10}$$

Subtracting (8) from (10), we obtain

$$D_i(t) = x_i(t) - (\Phi^t x(t))_i = \int_0^t (b_i(x(s)) - b_i(\Phi^t x(s)))ds + M_i(t), \quad i \in I. \tag{11}$$

We will view $M(t) = (M_i(t))_{i \in I}$ as a vector-valued martingale. To estimate the integral term, we recall the definition (3) and prove the following statement:

**Lemma 1.** *Let $F = \max_{i \in \mathbb{T}} f_i$. Then, for all $I \subset \mathbb{T}$,* $\quad \| b(x) - b(y) \| \leq 3F \| x - y \|, \quad x, y \in \Delta_I$.

Proof. We have

$$\| b(x) - b(y) \| = \sum_i |b_i(x) - b_i(y)| = \sum_i \left| \left( f_i x_i - x_i \sum_j x_j f_j \right) - \left( f_i y_i + y_i \sum_j y_j f_j \right) \right|$$

$$\leq J_1(x,y) + J_2(x,y),$$

where

$$J_1(x,y) = \left| \sum_i f_i (x_i - y_i) \right| \leq F \| x - y \|$$

and

$$J_2(x,y) \leq \sum_i \left| x_i \sum_j x_j f_j - y_i \sum_j y_j f_j \right|$$

$$\leq \sum_i \left| x_i \left( \sum_j x_j f_j - \sum_j y_j f_j \right) + (x_i - y_i) \sum_j y_j f_j \right|$$

$$\leq \sum_i x_i J_1(x,y) + \sum_i |x_i - y_i| F \leq J_1(x,y) + F \| x - y \|_1 \leq 2F \| x - y \|.$$

Combining three displays above, we complete the proof. □

Taking the absolute value in (11), then taking the sum over $i \in I$ and applying Lemma 1, we obtain

$$\| D(t) \| \leq 3F \int_0^t \| D(s) \| \, ds + M^*(t),$$

where $M^*(t) = \sup_{s \in [0,t]} \| M(s) \|$.

Using the Gronwall inequality, we obtain

$$\| D(t) \| \leq M^*(t) e^{3Ft}. \tag{12}$$

To estimate $M^*(t)$, we first use (4) to write for any $\beta > 0$:

$$\mathsf{P}\{M^*(t) \geq N^{-\beta}\} \leq \sum_i \mathsf{P}\{M_i^*(t) \geq N^{-\beta-\mu}\} \leq N^\mu \max_{i \in I} \mathsf{P}\{M_i^*(t) \geq N^{-\beta-\mu}\}, \tag{13}$$

where $M_i^*(t) = \sup_{s \in [0,t]} |M_i(s)|$. Next, we will apply an exponential martingale inequality from (51)(Appendix B6) in the form given by van de Geer (52)(Lemma 2.1):

**Lemma 2.** *If jumps of a locally square integrable cadlag martingale $(M(t))_{t \geq 0}$ are uniformly bounded by a constant $K > 0$, then*

$$\mathsf{P}\{\exists t : |M(t)| \geq A, \langle M \rangle_t \leq B^2\} \leq 2\exp\left[ -\frac{A^2}{2(AK + B^2)} \right].$$

Each $M_i$ is a piece-wise linear martingale with jumps of size $1/N$ (its jumps coincide with those of $x_i(t)$). Since, in addition, the total jump rate is bounded by $NF$, we obtain that the predictable quadratic variation of $M_i$ satisfies $\langle M_i \rangle_t \leq tNF/N^2 = tF/N$. Thus, we can apply Lemma 2 with $B^2 = tF/N$, $K = 1/N$, and $A = N^{-\beta-\mu}$:

$$P\{M_i^*(t) \geq N^{-\beta-\mu}\} \leq 2\exp[-\frac{N^{-2(\beta+\mu)}}{2(N^{-(\beta+\mu)-1} + tFN^{-1})}], \quad i \in I.$$

Combining this with (13), choosing $\beta$ so that $\beta + \mu < 1/2$ and using $t = c\ln N$, we can find constants $C, \gamma > 0$ such that

$$P\{M^*(t) \geq N^{-\beta}\} \leq 2N^\mu\exp[-\frac{N^{-2(\beta+\mu)}}{2(N^{-(\beta+\mu)-1} + tFN^{-1})}] \leq Ce^{-N^\gamma}$$

Using this in (12), we complete the proof of Part 1 of the theorem. To prove Part 2, we notice that according to Part 1, up to a SE-unlikely event, the stochastic process follows the deterministic trajectory $N^{-\beta}$-closely up to time $\tau_e \wedge c\ln N$, where $\tau_e$ is the first moment when one of the types goes extinct. We can restart the process at $\tau_e \wedge c\ln N$ treating $x(\tau_e \wedge c\ln N)$ as a new starting point and apply the same estimate to the restarted process (in case $\tau_e < c\ln N$, with fewer nonzero coordinates involved). Patching several ODE trajectories together in this way and noting that, conditioned on nonextinction of type $i^*$, the total time it takes to travel from any point $x \in \Delta_I$ with $x_{i^*} \geq N^{-1}$ to the neighborhood of $e^{(i^*)}$ of size $N^{-\delta}$ is bounded by $C\ln N$ for some $C$, we obtain Part 2.

The remaining parts follow from an auxiliary statement. To state it, we define a jump Markov process $y(t)$ with values in $\{0, N^{-1}, 2N^{-1} ..., 1\}$ such that $y(0) = x(0)$ and $y(t)$ makes a jump from $x$ to $x + N^{-1}$ with rate $Nf^*x(1-x)$ and to $x - N^{-1}$ with rate $N\tilde{f}x(1-x)$, where $\tilde{f} < f^*$ was defined in (5).

**Lemma 3.** *1. The process $y(t)$ is stochastically dominated by $x_{i^*}(t)$. 2. The process $y(t)$ considered only at times of jumps is an asymmetric random walk on $\{0, N^{-1}, 2N^{-1} ..., 1\}$ with absorption at $0$ and $N$ and probabilities of a step to the right and left being $p$ and $1 - p$ where $p \in (1/2, 1)$ solves*

$$\frac{p}{1-p} = \frac{f^*}{\tilde{f}}.$$

Proof. The coordinate $x_{i^*}$ jumps to the right with rate $Nf_ix_{i^*}(1-x_{i^*})$ and to the left with rate

$$Nx_{i^*} \sum_{j\neq i^*} f_j x_j \leq Nx_{i^*}\tilde{f} \sum_{j\neq i^*} x_j = N\tilde{f}x_{i^*}(1-x_{i^*}).$$

So, the jump rates to the left for both processes coincide and the jump rates to the right for process $y(t)$ do not exceed those for process $x_{i^*}(t)$, and Part 1 of the lemma follows. To prove Part 2, it suffices to note that the ratio of the jump right rate to the jump left rate for process $y(t)$ is equal to $f^*/\tilde{f}$ everywhere (except the absorbing points $0$ and $1$). $\square$

To prove Part 3, we can use this lemma and the fact that if $m \geq N/2$, then

$$N\frac{m}{N}\frac{m-N}{N} \geq \frac{1}{2}(m-N),$$

which implies that (except for an exponentially improbable event that $x_{i^*}$ hits level $N/2$ before 1), the time it takes for all non-$i^*$ types to die out is stochastically dominated by the extinction time for the linear birth-and-death process with birth rate $\lambda_k = Ak$ and death rate $\mu_k = Bk$ where $A = \tilde{f}/2 < B = f^*/2$. The probabilty $p_k(t)$ of extinction by time $t$ starting with $k$ individuals was probably first computed in (53). There is a misprint in formula (78) in (53) but one can use formula (68) of that paper (for generating functions) to obtain

$$p_k(t) = (\frac{Be^{(B-A)t} - B}{Be^{(B-A)t} - A})^k = (1 - \frac{B-A}{Be^{(B-A)t} - A})^k.$$

Plugging $t = C'\ln N$ and $k = N^{1-\delta}$ into this formula we obtain

$$1 - p_{N^{1-\delta}}(C'\ln N) \quad = 1 - (1 - \frac{B-A}{BN^{C'(B-A)} - A})^{N^{1-\delta}}$$

$$\sim \frac{(B-A)N^{1-\delta}}{BN^{C'(B-A)} - A} \sim \frac{B-A}{B}N^{1-\delta-C'(B-A)},$$

and since $\alpha = C'(B-A) - 1 + \delta > 0$ if we choose $C'$ large enough, the desired result follows.

The last two parts of Theorem 1 follow from Lemma 3, and similar well-known statements for asymmetric random walks. $\square$

Figure 1

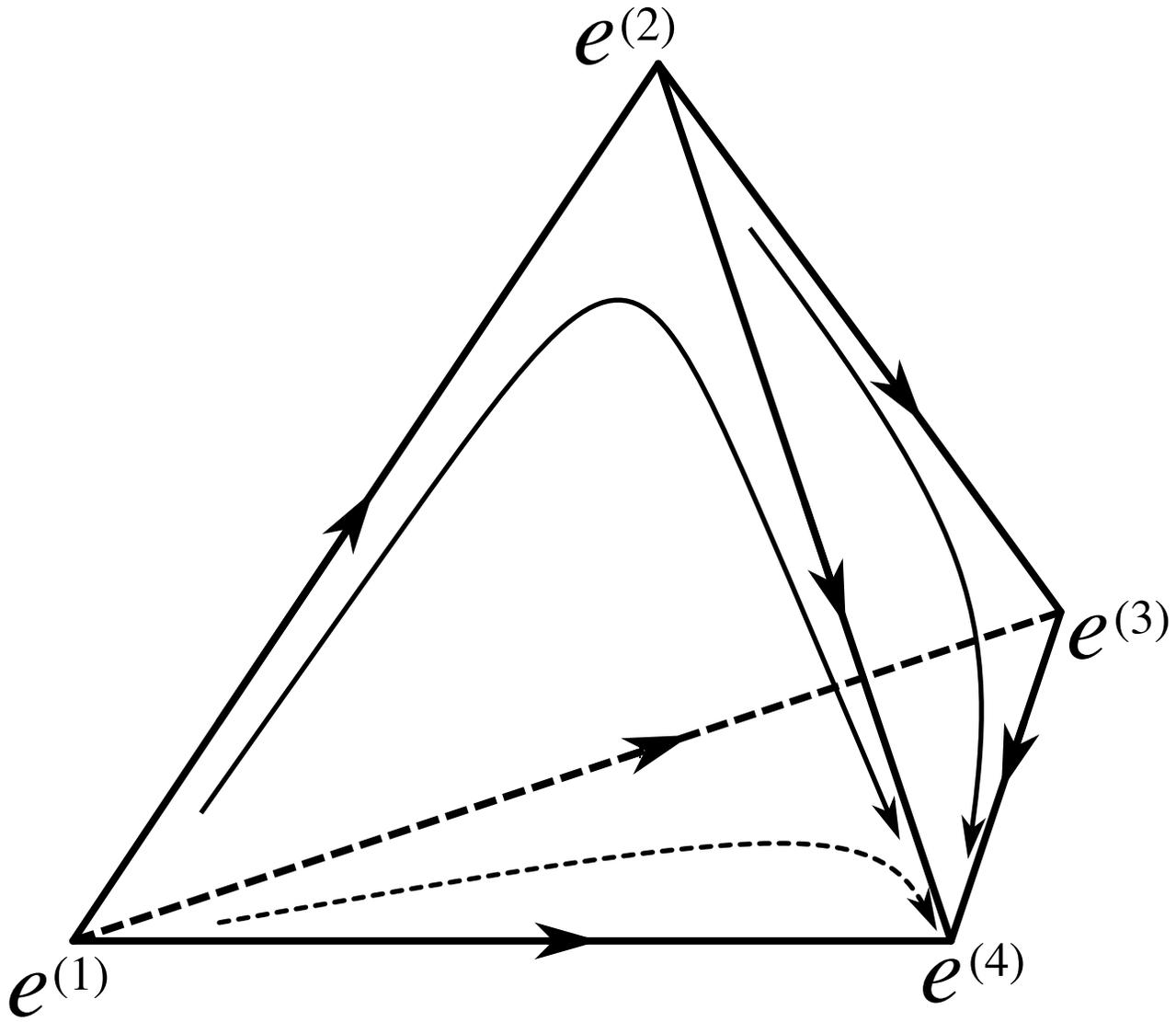

# Figure 2

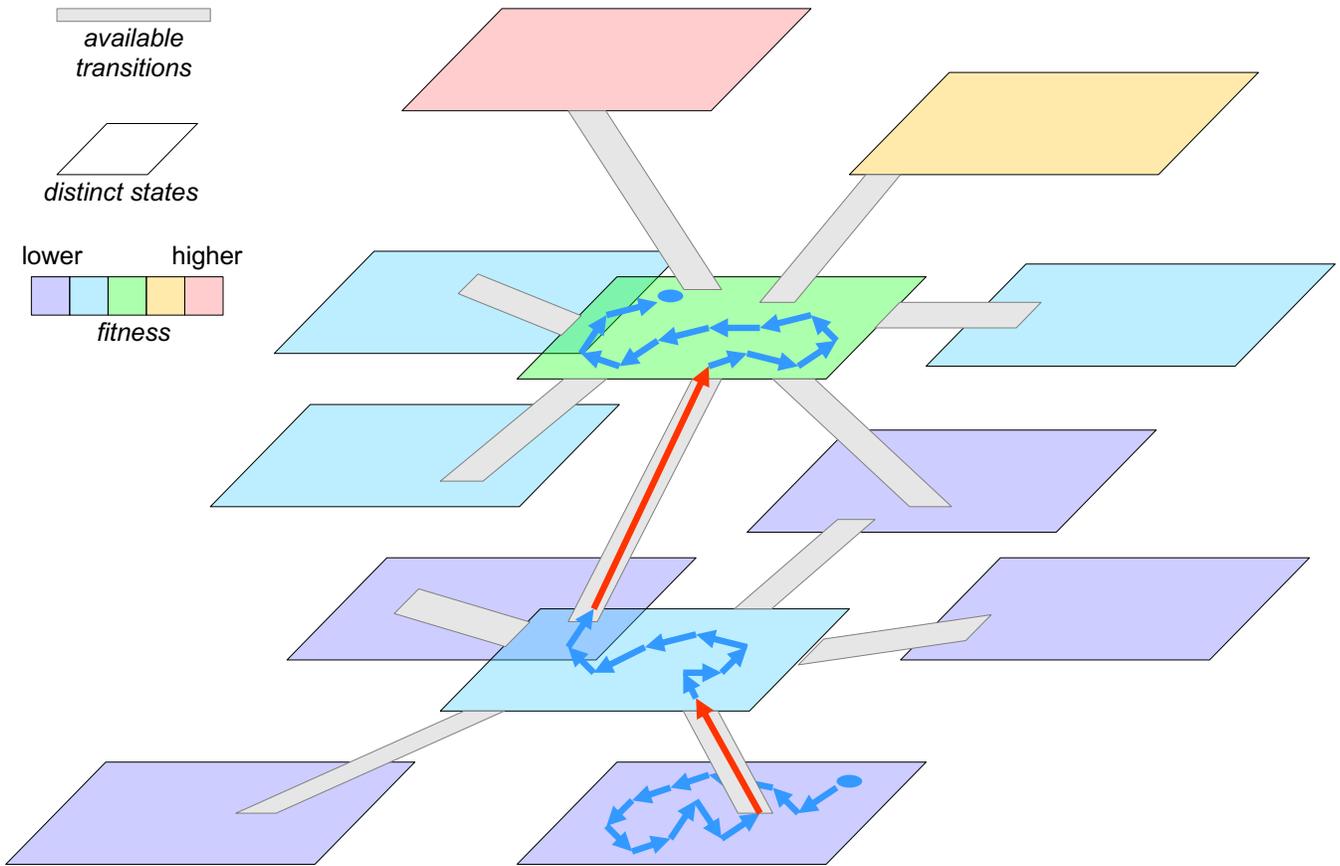